# Micromagnetic textures in exchange - coupled ferromagnetic – multiferroic films

Z.V. Gareeva[1,2], N.V. Shulga[1], A.K. Zvezdin[3,4]

[1]Institute of Molecule and Crystal Physics, Subdivision of the Ufa Federal Research Centre of the Russian Academy of Sciences, 450075, Ufa, Russia

[2]Physico-Technical Institute, Ufa University of Science and Technology, 450076, Ufa, Russia

[3]Prokhorov General Physics Institute of the Russian Academy of Sciences, 119991, Moscow, Russia

[4]"New spintronic technologies" Limited Liability Company, 121205, Skolkovo, Moscow, Russia

**Abstract**

The development of new computing technologies has given a new stimulus in the study of multiferroics. The use of multiferroics allows the realization of competitive energy efficient scalable logic and storage devices. The low - power consumption in Magneto Electric – Spin Orbital logics [1] and Magnetic Random Access Memory components is provided by magnetoelectric switching in multiferroic - based systems using a low-energy electric field. Our work concerns the modelling of the Magneto Electric – Spin Orbital elements with an emphasis on the magnetoelectric component and simulation of magnetization reversal processes in a model system. The use of the proposed approach makes it possible to analyze the influence of dimensional factors (film thicknesses, transverse dimensions, sample shape) affecting the magnetic states of multiferroic nanoelements; taking into interfacial interactions (magnetic anisotropy and interlayer exchange); energy-efficient external influences that allow switching magnetic states using magnetic and electric fields.

Corresponding authors: zukhragzv@yandex.ru, zvezdin@gmail.com

# 1. Introduction

The dramatic increase in the amount of big data used in various modern applications such as artificial intelligence, automotive driving, the internet of things stimulated progress in developing of data storage and processing devices. Although several competing computer architectures are currently been elaborated, the development of memory technologies, combined with advances in spintronics, is seen as a promising route in this direction [2,3]. To date, the process of creating commercial magnetic random access memory (MRAM) has gone through several stages: Toggle-MRAM (the first generation of commercial MRAM ); spin – transfer – torque (STT) MRAM (the second generation of commercial MRAM ); and spin-orbit-torque (SOT) MRAM under development, in which the read and write electric current paths to the magnetic – tunnel – junction element are separated from each other. Currently, both STT and SOT – MRAM are exploited in various processing – in-memory (PIM) architectures. The essence of PIM lies in the fact that the computations here are performed close to or inside memory blocks. In addition to the PIM direction, new spintonic logic devices (spin torque nanooscillators, spin-wave – based devices, Magnetoelectric Spin - Orbital (MESO) transistors) are being actively developed [1,4].

Herein, we refer to the MESO logics, the principal concepts of which are elaborated in details in Refs. [1]. Briefly, the basic MESO structure consists of three elements, the key one is nanomagnet connected with magnetoelectric and spin-orbit elements. Magnetization here is used as the computational variable. Compared with traditional logic devices, MESO logics has advantages such as power consumption saving and MESO circuits cascading. The most important parameters that characterize the efficiency of the device are (i) ultralow power consumption estimated as about 1aJ/switch; (ii) high values of magnetoelectric (ME) coefficient required to achieve effective ME coupling in the ME component of MESO; (iii) ultralow switching voltage to control ME effect, which is expected to be less than 100 mV; (iv) high efficiency of spin – to – charge transfer in the spin-orbital component characterized by the Hall angle of the 0.1 order; and (v) room operating temperatures [1, 2].

In this paper, we will focus on the ME component of the MESO device and consider multiferroics that are promising for its implementation, giving preference to compounds with room temperatures of ME orderings. ME effects in multiferroic heterostructures are discussed and a mathematical model of the magnetization reversal induced by an electric field in an exchange-coupled ferromagnetic-multiferroic (FM-MF) film is developed, as a candidate for the role of the magnetoelectric component of a MESO device.

# 2. Multiferroic heterostructures

Artificial multiferroic heterostructures begin to play a major role as relevant for technology. In multiferroic composites, the strength of the ME effect depends on the mechanisms that relate magnetic and ferroelectric (FE) orderings at the interface between magnetic and FE films. Basically, ME interface couplings can be realized through the strain effects and exchange interactions; current research also considers charge and ion transfer mechanisms [5]. Strains induce mechanical deformations of the piezoelectric and magnetic sublayers, which manifest themselves as magnetostriction in FM – FE film and exchange striction in antiferromagnetic (AFM) – FE films. The ionic mechanism involves controlling of magnetism of the FM layer through voltage-dependent chemical modification of the material, in particular through the migration of oxygen ions that affect magnetization. Charge transfer effects, leading to changes in the number of charge carriers in the magnetic layer, are more important for semi-metallic ferromagnets such as Heusler alloys. The ME response here is smaller compared to the response caused by strains.

Exchange interactions play an important role at the interfaces in FM-AFM and FM – MF heterostructures. Interactions between spins of AFM ordered MF and FM layer lead to (i) the bias effect attributed to the coupling of AFM ordering with spins in FM layer and (ii) exchange coupling between weak FM vector of MF with magnetization in FM. In the following consideration we take into account that (i) AFM vector aligns in such a way as to pin the spins in the FM layer in a certain direction dictated by the unidirectional magnetic anisotropy. (ii) Weak ferromagnetic vector is small enough and its interaction with magnetization of FM can be neglected. The exchange bias effect can be controlled using an electric field [6,7].

In this paper, we consider FM – MF bi-layer, taking the high temperature multiferroic $BiFeO_3$ as a MF component and a film with the parameters of iron garnets $R_3Fe_5O_{12}$ as FM component, and conduct calculations for the (110) – oriented $R_3Fe_5O_{12}$- $BiFeO_3$ film. Note that other MF and FM compounds may also be considered; for our model it is important to know their ground magnetic states, including information about the orientation of the AFM vector in MF film. To investigate the possible magnetic configurations in FM-MF film and their response to electric and magnetic fields we follow the approach described in details in Ref. [8]. We are just recapping the basic steps here. To account for the exchange coupling between MF with AFM ordering of spins and FM we introduce the pinning layer with strong unidirectional magnetic anisotropy. and divide the energy of the FM -MF structure into three parts

$$F = F_1(\boldsymbol{m}_1) + F_2(\boldsymbol{m}_2) + F_{MF}(\boldsymbol{P}, \boldsymbol{M}_{MF}, \boldsymbol{L}) \tag{1}$$

where

$$F_1(\boldsymbol{m}_1) = A_1(\nabla \boldsymbol{m}_1)^2 + K_{cub}\left(m_{1x}^2 m_{1y}^2 + m_{1y}^2 m_{1z}^2 + m_{1x}^2 m_{1z}^2\right) + K_{1u}\left(\boldsymbol{n}_u \boldsymbol{m}_1\right)^2 \tag{2}$$

is the energy of the soft FM (1) layer, where $A_1$ is the constant of exchange interactions, $K_{cub}$ is the constant of the cubic magnetic anisotropy, $K_{1u}$ is the constant of the uniaxial magnetic anisotropy ($K_{1u}$>0), $\boldsymbol{m}_1$ is the unit vector of magnetization in this layer

$$F_2(\boldsymbol{m}_2) = K_{2u}\left(\boldsymbol{n}_u \boldsymbol{m}_2\right)^2 - M_2\left(\boldsymbol{H} \cdot \boldsymbol{m}_2\right) \tag{3}$$

is the energy of the hard FM (2) layer, pinned to MF film, where $K_{2u}$ is the constant of the uniaxial magnetic anisotropy ($K_{2u}<0$), $\mathbf{m}_2$ is the unit vector of magnetization in the pinning layer, $\mathbf{H}$ is the magnetic field.

$$F_{MF}(\mathbf{M}_{MF},\mathbf{L},\mathbf{P}) = F_{me}(\mathbf{M}_{MF},\mathbf{L},\mathbf{P}) + F_{el}(\mathbf{P},\mathbf{E}) \qquad (4)$$

is the energy of MF divided into ferroelectric $F_{el}(\mathbf{P},\mathbf{E})$ and magnetoelectric $F_{me}(\mathbf{M}_{MF},\mathbf{L},\mathbf{P})$ parts. The ferroelectric energy $F_{el}(\mathbf{P},\mathbf{E})$ can be represented as an expansion in polarization components consistent with the symmetry of the MF under consideration; the magnetoelectric potential $F_{me}(\mathbf{M}_{MF},\mathbf{L},\mathbf{P})$ includes magnetic and magnetoelectric contributions dependent on ferromagnetic vector $\mathbf{M}_{MF} = \dfrac{M_{MF,1} + M_{MF,2}}{2M_0}$, AFM vector $\mathbf{L} = \dfrac{M_{MF,1} - M_{MF,2}}{2M_0}$, and polarization $\mathbf{P}$, $M_0$ is the saturation magnetization of the sublattices. For the BiFeO$_3$ the expressions for $F_{me}(\mathbf{M}_{MF},\mathbf{L},\mathbf{P})$ and $F_{el}(\mathbf{P},\mathbf{E})$ are given in Ref. [9,14,15]. In the case of BiFeO$_3$ $F_{me}(\mathbf{M}_{MF},\mathbf{L},\mathbf{P}) \ll F_{el}(\mathbf{P},\mathbf{E})$.

To find possible magnetic textures in the top FM film, we minimize energy (1). At first, using eq. (4), we obtain the equilibrium positions of AFM ($\mathbf{L}$) and weak FM ($\mathbf{M}_{MF}$) vectors in MF film at the chosen directions of applied electric field. We assume that the easy axis of uniaxial magnetic anisotropy in the pinning FM sublayer is determined by the direction of the AFM vector $\mathbf{L}$ in the MF film. After this, we conduct micromagnetic simulations of FM film, consisting of exchange - coupled soft (FM(1)) and hard (FM(2)) sublayers. This procedure makes it possible to determine the equilibrium distribution of magnetization in an FM film depending on the direction of the electric field.

### 3. Topological magnetic configurations in the FM-MF heterostructures

In this section, we present the results on simulation of micromagnetic states at definite orientation of the AFM vector coupled with polarization in MF film. Experiments on the electric field - induced switching in BiFeO$_3$ have shown that ferroelectric polarization can be switched by $71^0$, $109^0$ and $180^0$ by an out – of – plane electric field. This field – induced polarization $\mathbf{P}$ switching is accompanied with the rotation of AFM vector $\mathbf{L}$. In consistence with data given in Refs. [10,11], we chose orientation of $\mathbf{L}$ that corresponds to the definite orientation of $\mathbf{P}$ allocated by $\mathbf{E}$-field. To determine the magnetic state in FM film we conduct simulations using the Object Oriented Micro Magnetic Framework (OOMMF) public code [12]. Nanocell is taken of the dimensions $a \times a \times t$ nm$^3$, mesh size is $5 \times 5 \times 3$ nm$^3$, $a$=200 nm, $t$=50 nm, where $t$ is the thickness of FM layer, $a$ is the lateral width. As the pinning FM(2) layer, we take a layer of thickness 1 nm with strong magnetic anisotropy $K_{2u}$=1·10$^7$ J/m$^3$, the magnetic parameters of the FM(1) layer are taken close to the parameters of ferrite garnets: $A_1$=3·10$^{-12}$ J/m, $K_{cub}$=1·10$^4$ J/m$^3$, $K_{1u}$= −1·10$^4$ J/m$^3$ [13].

Magnetic textures in the FM sublayer is stabilized due to interplay of the Heisenberg exchange, magnetic anisotropy and magnetostatic interactions. In our previous study [8] we investigated the quasi-homogeneous magnetic configurations realized in the FM-MF nanoelement, here we focus on topological magnetic states. Inomogeneous distributions of magnetic moments in the (110) – oriented film are shown in Table 5. The initial positions of the polarization and AFM vector are given in the 2$^{nd}$ column, the initial 2D magnetic configurations on the top of FM layer are shown in the 4$^{th}$ column; the final positions of the polarization and AFM vector are given in the 3$^{rd}$ column, the final 2D magnetic configurations on the top of FM layer are shown in the 4$^{th}$ column, 1$^{st}$ column contains the information on the directions of electric field, the angles between the initial and final orientation so AFM vectors for each considered case numbered there. As seen from Table 5, the switching of the electric polarization in the MF sublayer induces the reorientation of the magnetization in the FM film; which means that by choosing the appropriate direction of the electric field, one can obtains the required distribution of magnetization in the system.

Table 5. Magnetic textures in the FM-MF nanoelement at different orientations of $E$ – field. Letters $L_0$ and $P_0$ stand for the initial directions of AFM and polarization vectors, $M_0(x,y)$ stands for the initial projection of magnetic moments onto the (110) plane in the upper layer of the FM film; $L_f$ and $P_f$ stand for the final directions of AFM and polarization vectors, $M_f(x,y)$ stands for the final projection of magnetic moments onto the (110) plane in the upper layer of the FM film, realized after E-field switching.

| № | $L_0$ | $L_f$ | $M_0$ | $M_f$ |
|---|---|---|---|---|
| Compressive deformations | | | | |
| 1<br>$E[\bar{1}\bar{1}0]$ | $L_0[11\bar{2}]$<br>$P_0[111]$ | $L_f[112]$<br>$P_f[\bar{1}\bar{1}1]$ | | |
| 2<br>$E[00\bar{1}]$ | $L_0[11\bar{2}]$<br>$P_0[111]$ | $L_f[\bar{1}\bar{1}2]$<br>$P_f[11\bar{1}]$ | | |
| Tensile deformation | | | | |
| 3<br>$E[00\bar{1}]$ | $L_0[\bar{1}1\bar{2}]$<br>$P_0[\bar{1}11]$ | $L_f[1\bar{1}2]$<br>$P_f[\bar{1}1\bar{1}]$ | | |
| 4<br>$E[1\bar{1}0]$ | $L_0[1\bar{1}2]$<br>$P_0[\bar{1}11]$ | $L_f[\bar{1}12]$<br>$P_f[1\bar{1}1]$ | | |

## 4. Conclusion

Multiferroic heterostructures that can be used in MESO- devices are analyzed. The ME properties in multiferroic composites are compared and the problem of modeling the magnetization reversal process in an exchange - coupled FM-MF bilayer is considered. A model is proposed that makes it possible to combine the calculation of the ground magnetic states of the ferromagnetic and multiferroic subsystems due to the bias effect. Possible magnetic textures in a (110) – oriented nanoscale $R_2Fe_5O_{12}$-$BiFeO_3$ film, which appear under the action of an electric field acting on the MF sublayer, are calculated. In the framework of the proposed approach, it is further possible to study problems that are interesting for practical applications: analysis of the influence of dimensional factors, such as film thickness and sample shape; topological magnetic states; pinning layers with different couplings, including interfacial Dzyaloshinskii – Moriya interaction on magnetization switching induced by electric field.

## Acknowledgement

Z.V.Gareeva acknowledges the Support by the State assignment for the implementation of scientific research by laboratories (Order MN-8/1356 of 09/20/2021), N.V. Shulga acknowledges the Support by Russian Science Foundation №23-22-00225, A.K. Zvezdin acknowledges the Support by Ministry of Science and Higher Education of the Russian Federation, Agreement № 075-11-2022-046.